\documentclass[12pt]{article}
\usepackage{geometry}
\usepackage{here}
\usepackage{graphicx}
\usepackage{amsmath,amssymb}
\usepackage{cite}
\usepackage{bm}
\newcommand{\lw}[1]{\smash{\lower 1.5ex\hbox{#1}}}
\newcommand{\ri}[1]{\smash{\raise 1.5ex\hbox{#1}}}
\newcommand{\riw}[1]{\smash{\raise 3.0ex\hbox{#1}}}

\newcommand{\mapright}[1]{\smash{\mathop{\hbox to 3.0cm{\rightarrowfill}}\limits^{\displaystyle #1}}}

\begin{document}

\title{On long-range pionic Bose-Einstein correlations\\
-- Including analyses of OPAL, L3 and CMS BECs --}

\author{Takuya Mizoguchi$^{1}$ and Minoru Biyajima$^{2}$\\
{\small $^{1}$National Institute of Technology, Toba College, Toba 517-8501, Japan}\\
{\small $^{2}$Department of Physics, Shinshu University, Matsumoto 390-8621, Japan}}

\maketitle

\begin{abstract}
Long-range correlation plays an important role in analyses of pionic Bose-Einstein correlations (BECs). In many cases, such correlations are phenomenologically introduced. In this investigation, we propose an analytic form. By making use of the form, we analyze the OPAL BEC and the L3 BEC at $Z^0$-pole and the CMS BEC at 0.9 and 7 TeV using our formulas and the $\tau$-model. The parameters estimated by both approaches are found to be consistent. Utilizing the Fourier transform in four-dimensional Euclidean space, a number of pion-pair density distributions are also studied.
\end{abstract}


\section{\label{sec1}Introduction}
The following conventional formula is utilized as a standard tool in many analyses of pionic Bose-Einstein correlation (BEC)~\cite{Acton:1991xb,Abreu:1992gj,Achard:2011zza,Aad:2015sja,Khachatryan:2011hi,Aaij:2017oqu,Biyajima:2018abe,Mizoguchi:2019cra,Biyajima:2019wcb,Mizoguchi:2020}:
\begin{eqnarray}
{\rm CF_{I}} = \left[1.0 + \lambda E_{\rm BE}(R,\,Q)\right]\cdot {\rm LRC},
\label{eq1}
\end{eqnarray}
where $E_{\rm BE}$ is the exchange function between two identical pions and $Q=-\sqrt{-(p_1-p_2)^2}$ is the magnitude of the momentum transition squared between them. Typically, $E_{\rm BE}$ is given by the Gaussian distribution and/or the exponential function. The degree of coherence is expressed by $\lambda$. The following long-range correlation (LRC) is frequently used:
\begin{eqnarray}
{\rm LRC}_{(\delta)}=C(1+\delta Q)
\label{eq2}
\end{eqnarray}
In this present paper, we first pay attention to the BEC created at the Z$^0$-pole by the OPAL collaboration, because they used a second kind of LRC, which was given as
\begin{eqnarray}
{\rm LRC}_{(\delta,\, \varepsilon)}=C(1+\delta Q+\varepsilon Q^2)
\label{eq3}
\end{eqnarray}
%

\begin{table}[h]
\centering
\caption{\label{tab1}Analysis of OPAL data using Eqs.~(\ref{eq1})--(\ref{eq2}). ``G'' denotes the Gaussian distribution.}
\vspace{1mm}
\begin{tabular}{ccccccc}
\hline
LRC & $R$ (fm)(G) & $\lambda$ & $c$ & $\delta$ (GeV$^{-1}$) & $\varepsilon$ (GeV$^{-2}$) & $\chi^2/$ndf\\
\hline
($\delta$)
& 1.12$\pm$0.03
& 0.78$\pm$0.04
& 0.73$\pm$0.00
& 0.16$\pm$0.00
& --- 
& 259/60\\

($\delta$, $\varepsilon$)
& 0.95$\pm$0.02
& 0.88$\pm$0.04
& 0.63$\pm$0.01
& 0.50$\pm$0.04
& $-$0.13$\pm$0.01
& 113/59\\
\hline
\end{tabular}
\end{table}

\begin{figure}[htbp]
  \centering
  \includegraphics[width=0.48\columnwidth]{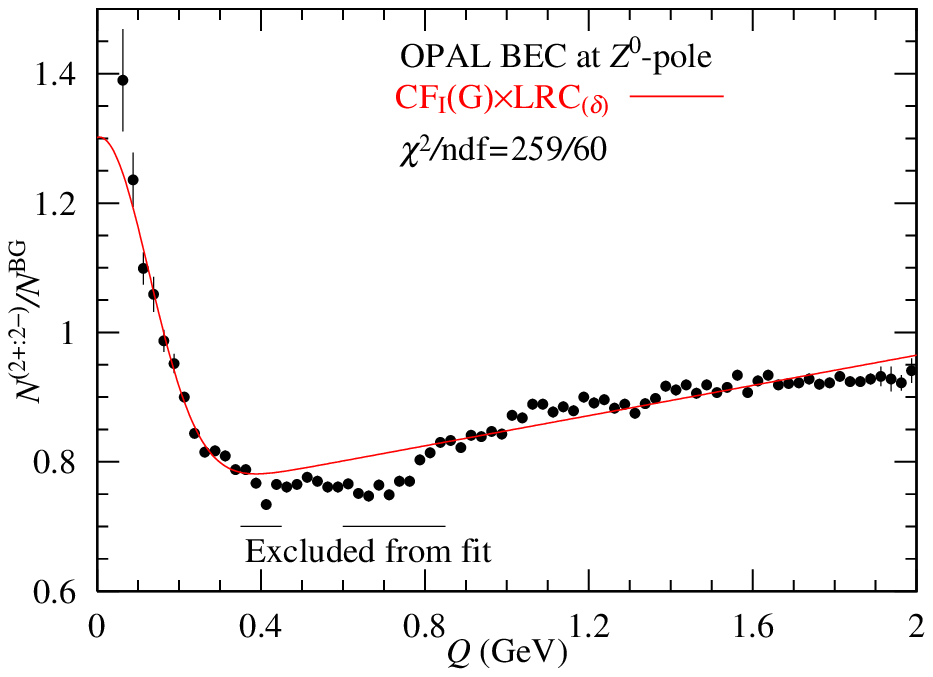}
  \includegraphics[width=0.48\columnwidth]{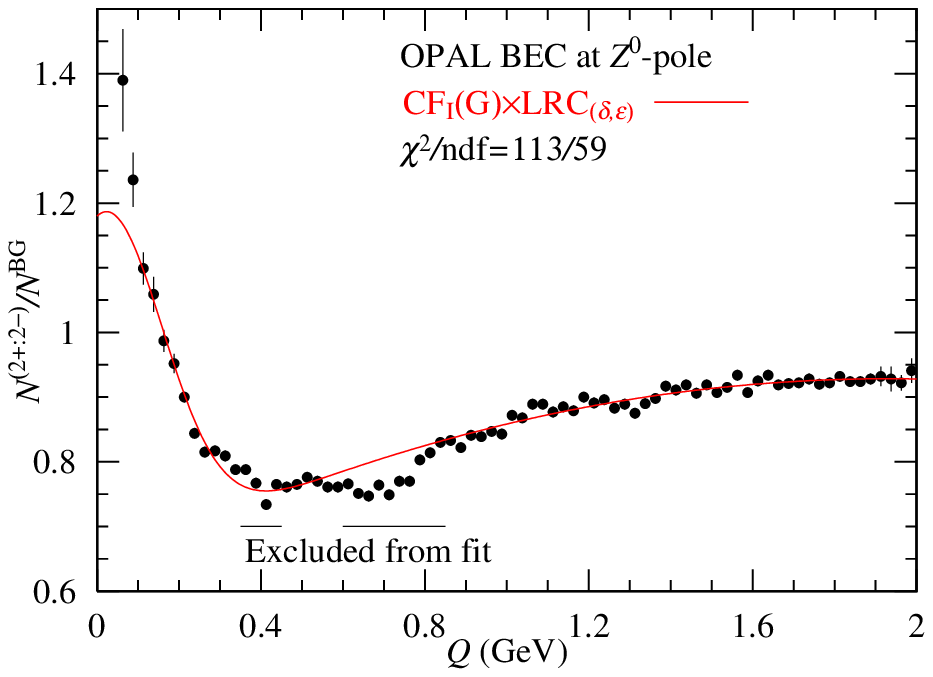}
  \caption{\label{fig1}Analysis of the OPAL BEC at the $Z^0$-pole using by Eq.~(\ref{eq1}) with Eqs.~(\ref{eq2}) and (\ref{eq3}). }
\end{figure}

As is seen in Table~\ref{tab1} and Fig.~\ref{fig1}, when LRC$_{(\delta,\, \varepsilon)}$ (i.e., Eq.~(\ref{eq3})) is utilized in the analysis, we obtain a better $\chi^2$ value than that using Eq.~(\ref{eq2}). This fact may suggest us that Eq.~(\ref{eq3}) reflects some physical meaning.\\

In Section~\ref{sec2}, we consider the analytic form of the LRC. In Section~\ref{sec3}, we analyze the CMS BEC created at 0.9 and 7 TeV by the CMS Collaboration using Eqs.~(\ref{eq3}) and (\ref{eq4}) and an analytic form mentioned in the next section. In Section~\ref{eq4}, we analyze the OPAL BEC at the $Z^0$-pole using a new conventional formula introduced in Section~\ref{sec3}. In Section~\ref{sec5}, by making use of the Fourier transform, we show the density distribution of pion-pairs in a four-dimensional Euclidean space, where $\xi = (x^2+y^2+z^2+(ct)^2)^{1/2}$ is introduced. Finally, in Section~\ref{sec6}, concluding remarks and discussions are presented.


\section{\label{sec2}Long-range Correlation}
It is remarkable that Eq.~(\ref{eq3}) utilized by the OPAL collaboration works well; given that it is a phenomenological form that depends upon the parameters $\delta$ and $\varepsilon$, its asymptotic behavior is as follows:
$$
\mbox{Eq.~(\ref{eq3})}\mapright{Q\to \infty} -\mbox{large value}.
$$
In Table~\ref{tab1}, at $Q=0.0$ the normalization factor $c$ is $0.65\pm 0.01$, which is much lower than $C\cong 1.0$. To avoid this behavior at $Q\to \infty$, we propose the following analytic form: 
\begin{eqnarray}
{\rm LRC}_{(\alpha,\, \beta,\, n)}=C[1+\alpha Q^n\exp(-\beta Q)],
\label{eq4}
\end{eqnarray}
where $\alpha$ and $\beta$ are parameters. As $n=1$,
\begin{eqnarray}
\mbox{Eq.~(\ref{eq4})}\left\{
\begin{array}{l}
\mapright{Q\to \infty} C\simeq 1.0,\medskip\\
\mapright{Q\to 0} C(1.0+\alpha Q - \alpha \beta Q^2).
\end{array}
\right.
\label{eq5}
\end{eqnarray}
This is the same form as Eq.~(\ref{eq3}); moreover, to make certain a case with $n=2$ is investigated. In addition to Eq.~(\ref{eq5}), the following correspondences are expected:
\begin{eqnarray*}
\left\{
\begin{array}{l}
\alpha > 0\ \cdots\ \mbox{reproducing Eqs.~(\ref{eq2}) and (\ref{eq3})},\medskip\\
\alpha \approx 0\ \cdots\ \mbox{no any effect},\medskip\\
\alpha < 0\ \cdots\ \mbox{because of negative contribution to BEC, this sign implies the subtraction}\\
\qquad\qquad\mbox{of non-BE effect: the contamination between different hadron pairs, resonances}\\
\qquad\qquad\mbox{effect, and/or the energy conservation.}
\end{array}
\right.
\end{eqnarray*}

By using Eq.~(\ref{eq4}) with smaller $\chi^2$ values, we are able to determine some physical information contained within the LRCs. 


\subsection{Analysis of the OPAL BEC data}
\begin{figure}[h]
  \centering
  \includegraphics[width=0.48\columnwidth]{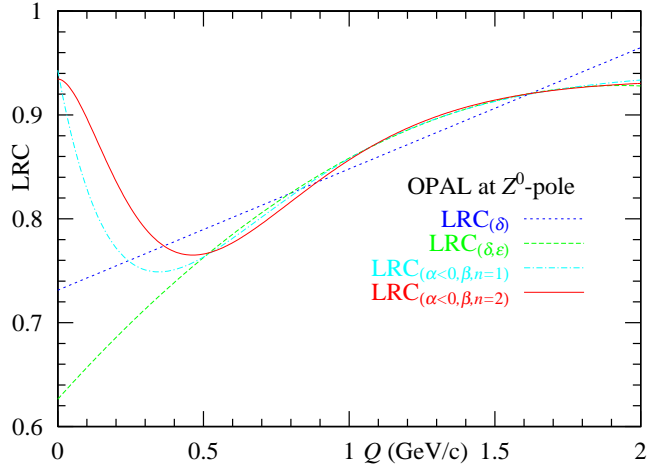}
  \caption{\label{fig2}LRC in the OPAL BEC at the $Z^0$-pole ($\alpha < 0$) (see Tables~\ref{tab1} and \ref{tab2}).}
\end{figure}

It should be noted that the OPAL collaboration reported two kinds of data, i.e., an ``ordinary'' data ensemble and a ``corrected'' data ensemble that had been renormalized using the Monte Carlo calculation
\begin{eqnarray*}
\left\{
\begin{array}{l}
N^{(2+:\,2-)}/N^{\rm BG},\medskip\\
N_{\rm MC}^{(2+:\,2-)}/N_{\rm MC}^{\rm BG} = \dfrac{N^{(2+:\,2-)}/N_{\rm MC}^{(2+:\,2-)}}{N^{BG}/N_{\rm MC}^{\rm BG}}.
\end{array}
\right.
\end{eqnarray*}
\paragraph{(a)} First of all, we analyze the OPAL BEC at the $Z$-pole. Our results by means of Eq.~(\ref{eq4}) with $n=1$ and 2 are displayed in Table~\ref{tab2} and Fig.~\ref{fig2}. As $\alpha>0$ in Eq.~(\ref{eq4}), the two cases are similar to those using LRC$_{(\delta,\, \varepsilon)}$ in Table~\ref{tab1}. For $\alpha<0$, improvements about $\chi^2$'s are seen in Table~\ref{tab2}. This fact probably means that the OPAL BEC at the $Z^0$-pole contains non-BE effects, which must be subtracted from the data (see Section~\ref{sec6}). 

\begin{table}[h]
\centering
\caption{\label{tab2}Analysis of the OPAL data using Eqs.~(\ref{eq1}) and (\ref{eq4}). Values set with *) indicate that $|\alpha|=6.24$ GeV$^{-2}$ is larger than $|\alpha|=1.61$ GeV$^{-1}$ for $n=1$.}
\vspace{1mm}

\begin{tabular}{ccccccc}
\hline
LRC & $R$ (fm)(G) & $\lambda$ & $c$ & $\alpha$ (GeV$^{-n}$ ) & $\beta$ (GeV$^{-1}$) & $\chi^2/$ndf\\
\hline
$\alpha>0$\\
($n=1.0$)
& 0.94$\pm$0.02
& 0.93$\pm$0.04
& 0.60$\pm$0.01
& 0.66$\pm$0.07
& 0.44$\pm$0.02
& 120/59\\

($n=2.0$)
& 0.96$\pm$0.03
& 0.77$\pm$0.03
& 0.70$\pm$0.01
& 0.65$\pm$0.06
& 1.03$\pm$0.03
& 104/59\\
\hline
\hline

$\alpha<0$\\
($n=1.0$)
& 0.94$\pm$0.05
& 0.48$\pm$0.03
& 0.94$\pm$0.00
& $-$1.61$\pm$0.09
& 2.89$\pm$0.11
& 91.3/59\\

($n=2.0$)
& 1.25$\pm$0.06
& 0.46$\pm$0.04
& 0.93$\pm$0.00
& $-$6.24$\pm$0.27
& 4.32$\pm$0.08
& 77.5/59 *)\\
\hline
\end{tabular}
\end{table}

\begin{figure}[h]
  \centering
  \includegraphics[width=0.48\columnwidth]{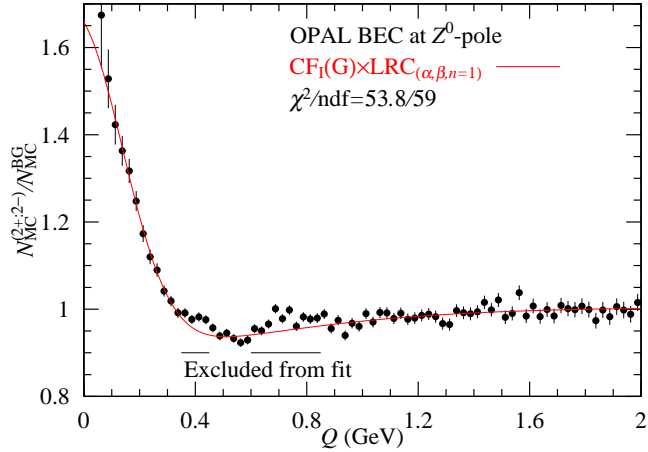}
  \caption{\label{fig3}Analysis of the OPAL BEC at the $Z^0$-pole renormalized using the Monte Carlo calculation based on Eqs.~(\ref{eq1}) and (\ref{eq4}). }
\end{figure}

\paragraph{(b)} Next, we analyze the second data ensemble by means of Eqs.~(\ref{eq1}) and (\ref{eq4}). Our results are shown in Fig.~\ref{fig3} and Table~\ref{tab3}.

%
\begin{table}[H]
\centering
\caption{\label{tab3}Analysis of the OPAL BEC at the $Z^0$-pole renormalized using the Monte Carlo calculation based on Eqs.~(\ref{eq1}) and (\ref{eq4}). }
\vspace{1mm}

\begin{tabular}{ccccccc}
\hline
LRC & $R$ (fm)(G) & $\lambda$ & $c$ & $\alpha$ (GeV$^{-n}$ ) & $\beta$ (GeV$^{-1}$) & $\chi^2/$ndf\\
\hline
$\alpha>0$\\
($n=1.0$)
& 0.91$\pm$0.03
& 0.74$\pm$0.04
& 0.90$\pm$0.02
& 0.14$\pm$0.05
& 0.45$\pm$0.12
& 112/73\\

($n=2.0$)
& 0.91$\pm$0.03
& 0.71$\pm$0.04
& 0.93$\pm$0.01
& 0.15$\pm$0.06
& 1.05$\pm$0.15
& 112/73\\

\hline
\hline

$\alpha<0$\\
($n=1.0$)
& 0.89$\pm$0.04
& 0.64$\pm$0.03
& 1.00$\pm$0.06
& $-$0.52$\pm$0.16
& 3.22$\pm$0.62
& 111/73\\

($n=2.0$)
& 0.92$\pm$0.04
& 0.61$\pm$0.04
& 1.00$\pm$0.00
& $-$2.08$\pm$0.65
& 4.76$\pm$0.60
& 112/73\\
\hline
\end{tabular}
\end{table}


\subsection{Analysis of the L3 BEC data}
\begin{figure}[htbp]
  \centering
  \includegraphics[width=0.48\columnwidth]{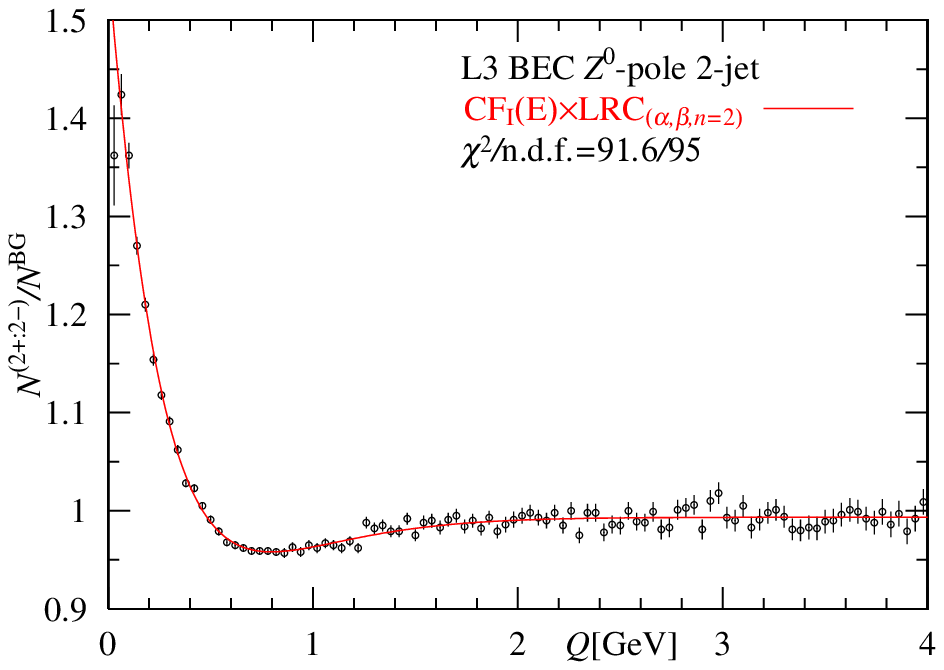}
  \includegraphics[width=0.48\columnwidth]{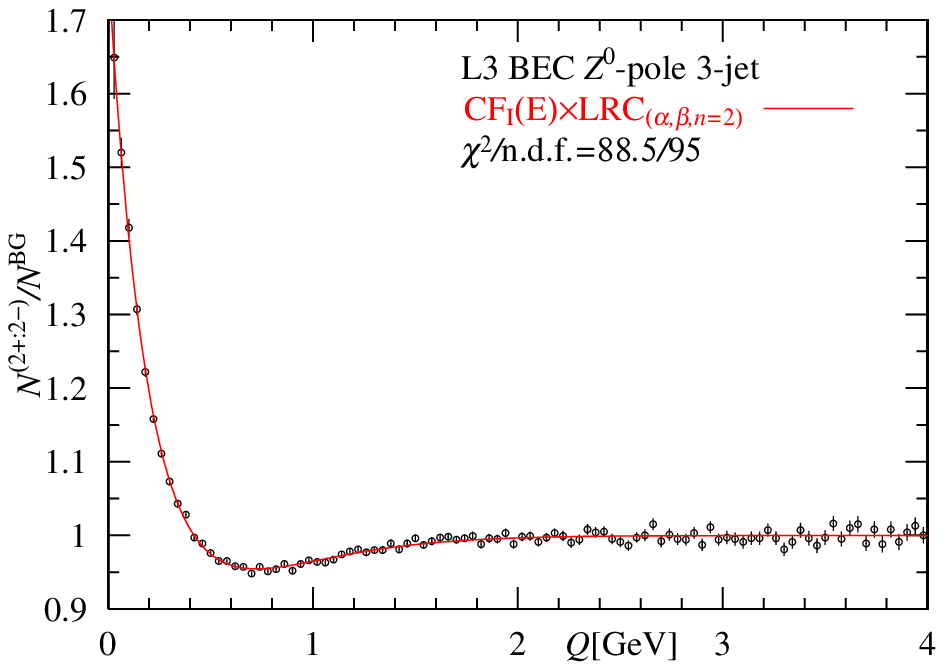}
  \caption{\label{fig4}Analysis of the L3 BEC at the $Z^0$-pole using Eqs.~(\ref{eq1}) and (\ref{eq4}). }
\end{figure}

\begin{figure}[H]
  \centering
  \includegraphics[width=0.48\columnwidth]{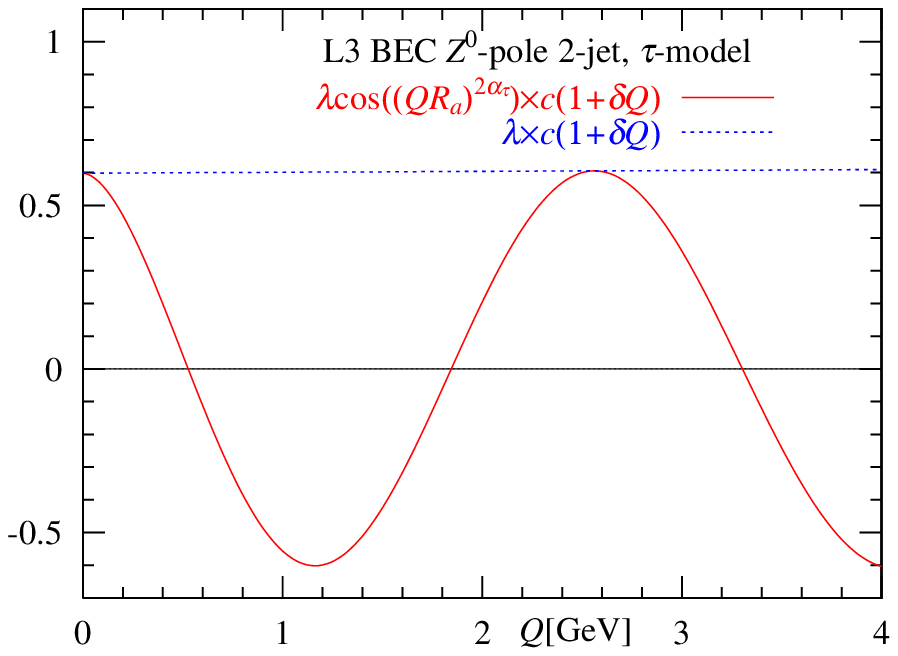}
  \includegraphics[width=0.48\columnwidth]{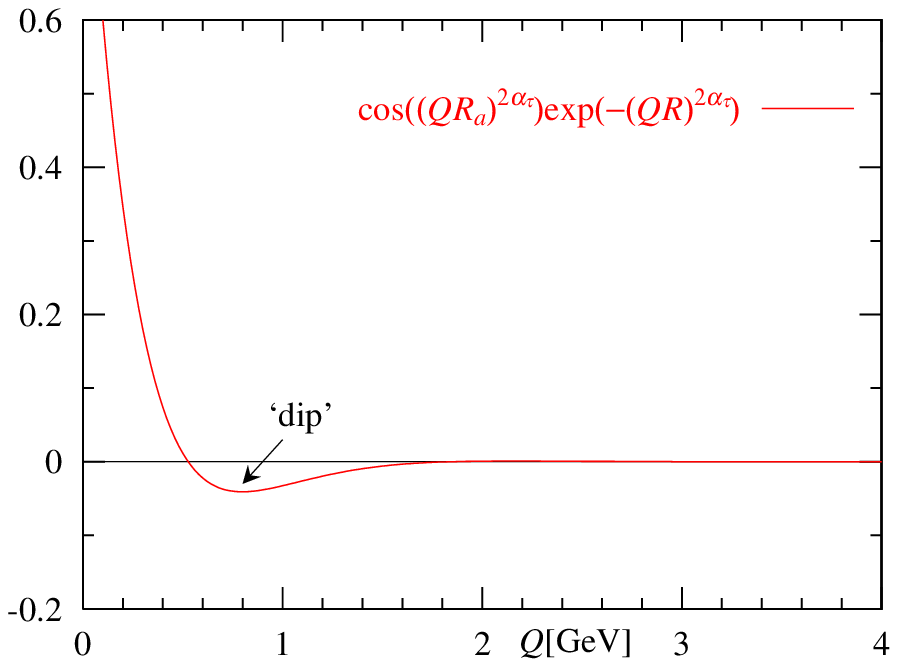}
  \caption{\label{fig5}``Effective degree of coherence'' and ``dip structure'' in the $\tau$-model.}
\end{figure}

Because the L3 collaboration reported BEC data for 2-jet ($q\bar q$ jet) and 3-jet ($q\bar qg$ jet) cases, we are interested in analyzing those data. Such data can be categorized into the same kinds of ensembles with two and 3three jets, respectively. Thus, we may analyze them using the ${\rm CF_{I}}$ with ${\rm LRC}_{(\alpha,\, \beta,\, n)}$. Our results are shown in Fig.~\ref{fig4} and Table~\ref{tab4}. To compare them with those obtained using the $\tau$-model~\cite{Csorgo:2003uv,Csorgo:2008ah,Zolotarev,Sato:1990,Sato:1999}, 
\begin{eqnarray}
F_{\tau}(e^+e^-) = \left\{1 + \lambda \cos \left[(R_aQ)^{2\alpha_{\tau}}\right] \exp\left[-(RQ)^{2\alpha_{\tau}}\right]\right\}\times {\rm LRC}_{(\delta)},
\label{eq6}
\end{eqnarray}
with $R_a^{2\alpha_{\tau}}=\tan(\alpha_{\tau}\pi/2)R^{2\alpha_{\tau}}$, we analyze them. As seen in Fig.~\ref{fig5}, the effective degree of coherence in the $\tau$-model is oscillating. The results from Eq.~(\ref{eq6}) are also presented in Table~\ref{tab4}. 

%
\begin{table}[H]
\centering
\caption{\label{tab4}Analysis of the L3 BEC at the $Z^0$-pole using Eqs.~(\ref{eq1}) and (\ref{eq4}) and the $\tau$-model.}
\vspace{1mm}
\begin{tabular}{ccccccc}
\hline
\multicolumn{3}{l}{${\rm CF_{I}}$ with ${\rm LRC}_{(\alpha,\, \beta,\, n=2)}$}\\
event & $R_1$ (fm)(E) & $\lambda$ & $c$ & $\alpha$ (GeV$^{-2}$ ) & $\beta$ (GeV$^{-1}$) & $\chi^2/$ndf\\
\hline
 2-jet
& 0.87$\pm$0.08
& 0.57$\pm$0.03
& 0.993$\pm$0.001
& $-$1.76$\pm$0.42
& 3.86$\pm$0.21
& 91.6/95\\

 3-jet
& 1.19$\pm$0.05
& 0.77$\pm$0.03
& 1.000$\pm$0.001
& $-$1.62$\pm$0.18
& 3.78$\pm$0.13
& 88.5/95\\
\hline
\hline
\multicolumn{3}{l}{$\tau$-model with ${\rm LRC}_{(\delta)}$}\\
event & $R$ (fm) & $\lambda$ & $c$ & $\alpha_{\tau}$ & $\delta$ (GeV$^{-1}$) & $\chi^2$/ndf\\
\hline
 2-jet
& 0.78$\pm$0.04
& 0.61$\pm$0.03
& 0.979$\pm$0.002
& 0.44$\pm$0.01
& 0.005$\pm$0.001
& 95/95\\

 3-jet
& 0.99$\pm$0.04
& 0.85$\pm$0.04
& 0.977$\pm$0.001
& 0.41$\pm$0.01
& 0.008$\pm$0.001
& 112/95\\

\hline
\end{tabular}
\end{table}

\section{\label{sec3}Analysis of CMS BEC}
\begin{figure}[H]
  \centering
  \includegraphics[width=0.48\columnwidth]{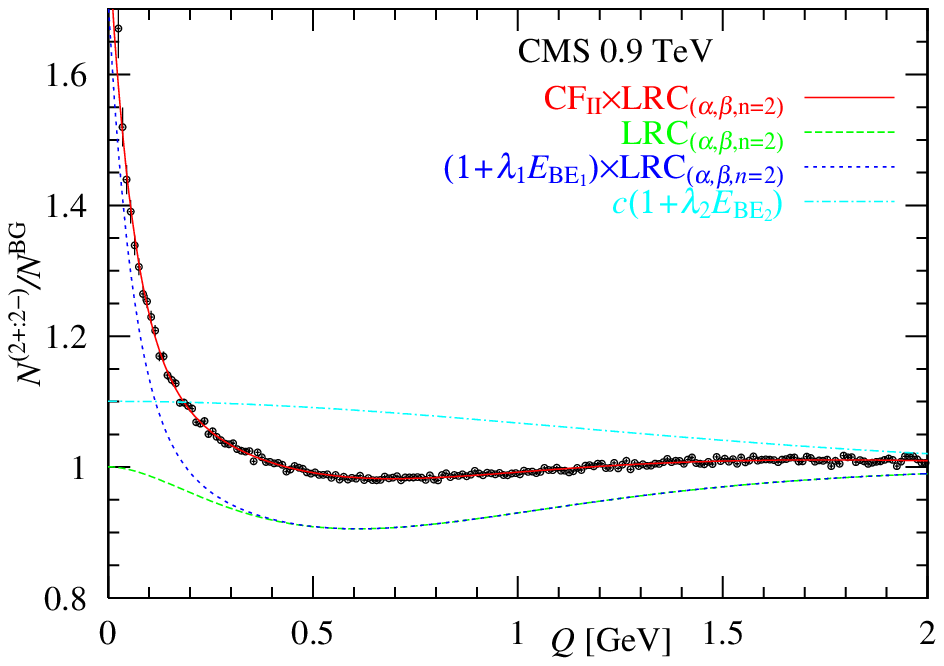}
  \includegraphics[width=0.48\columnwidth]{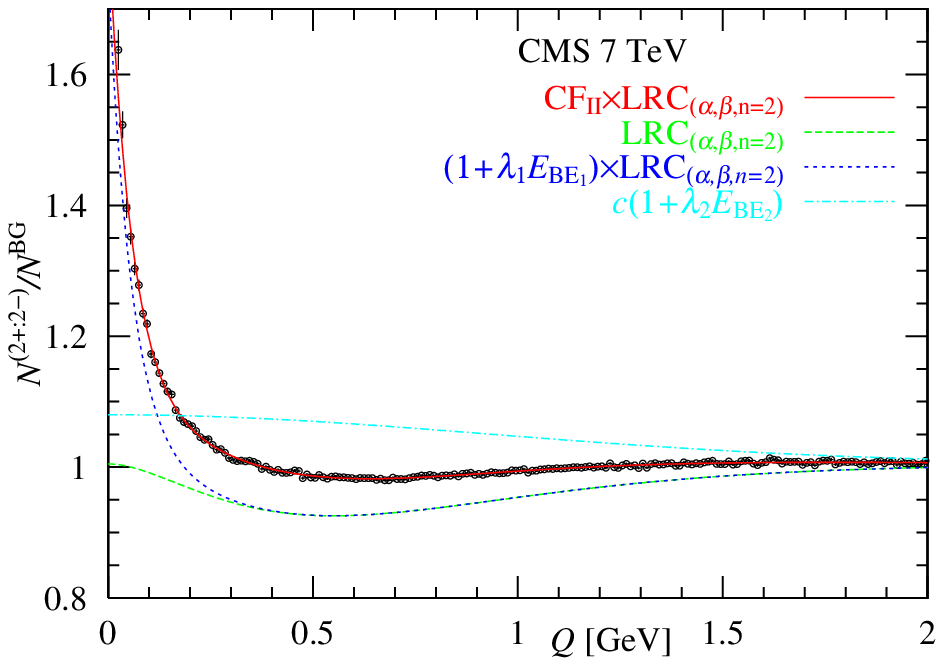}
  \caption{\label{fig6}Analysis of the CMS BEC at 0.9 and 7 TeV.}
\end{figure}

\begin{figure}[H]
  \centering
  \includegraphics[width=0.48\columnwidth]{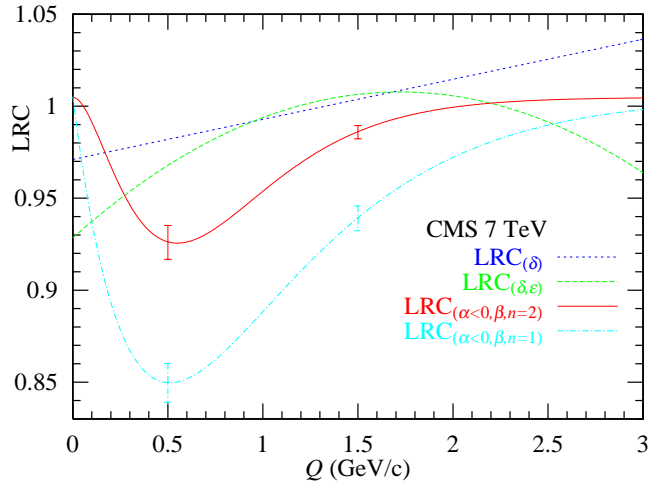}
  \caption{\label{fig7}LRC in the CMS BEC at 7 TeV ($\alpha<0$). The bar indicates the band of the LRCs.}
\end{figure}

We analyze the CMS BEC at 0.9 and 7 TeV using the following formula:
\begin{eqnarray}
{\rm CF_{II}} = \left[(1.0 + \lambda_1 E_{\rm BE_1}(R_1,\,Q) + \lambda_2 E_{\rm BE_2}(R_2,\,Q)\right]\cdot {\rm LRC},
\label{eq7}
\end{eqnarray}
where the second $\lambda$ ($\lambda_2$) and the second exchange function ($E_{\rm BE_2}$) are introduced to describe the BEC data at the LHC. A detailed derivation and analysis with Eq.~(\ref{eq3}) (i.e., ${\rm LRC}_{(\delta)}$) are presented in Refs.~\cite{Biyajima:2018abe,Biyajima:2019wcb,Mizoguchi:2020}. For our purposes, the analytic ${\rm LRC}_{(\delta,\, \varepsilon)}$ is also necessary in the CF$_{\rm II}$. Our results are shown in Figs. \ref{fig6} and \ref{fig7}, and Table~\ref{tab5}. 

In Table~\ref{tab5}, we also show the results obtained using the $\tau$-model formula, which is appropriate for LHC collisions. This means that there are two formulas, Eq.~(\ref{eq6}) for $e^+e^-$ collisions and Eq.~(\ref{eq8}) for LHC collisions. Thus, to analyze the CMS BEC at 0.9 and 7 TeV, the authors of \cite{Khachatryan:2011hi,Csorgo:2003uv,Csorgo:2008ah,Sirunyan:2019umv} employ the following equation:
\begin{eqnarray}
F_{\tau} = \left\{1.0 + \lambda \cos\left[(R_0Q)^2+\tan\left(\frac{\alpha_{\tau}\pi}4\right)(RQ)^{\alpha_{\tau}}\right]\exp(-(RQ)^{\alpha_{\tau}})
\right\}\times {\rm LRC}_{(\delta)},
\label{eq8}
\end{eqnarray}
where $R_0$ is the free parameter.

As seen in Table~\ref{tab5} and Fig.~\ref{fig7}, three LRCs are grouped together and the estimated parameters are almost the same.

The three columns in Table~\ref{tab5} indicate that (in the center column) is almost the same as the set obtain using the $\tau$-model, provided that LRC$_{(\alpha,\, \beta,\, n=2)}$ is adopted. Comparing the second set with the first one with  LRC$_{(\delta,\, \varepsilon)}$, we find that the estimated $R_1$s in the first set are somewhat smaller than $R_1$ those in the second set. The situation is the opposite for $R_2$. This is probably attributable to the sets' normalization factors, $c=0.91$ or 0.93. 

\begin{table}[h]
\centering
\caption{\label{tab5}The estimated parameters for the CMS BEC at 0.9 and 7 TeV.} 
\vspace{1mm}
\renewcommand{\arraystretch}{1.2}
\begin{tabular}{c|c|c}
\hline
  CF$_{\rm II}\times$LRC$_{(\delta,\ \varepsilon)}$
& CF$_{\rm II}\times$LRC$_{(\alpha,\ \beta,\ n=2)}$
& $F_{\tau}$ ($\tau$-model)\\
\hline
0.9 TeV\qquad $R$ (fm)\qquad & $R$ (fm) & $R$ (fm)\\
  $R_1 = 2.79$, $\lambda_1 =0.82$ (E)
& $R_1 = 3.07$, $\lambda_1 = 0.71$ (E)
& $R = 2.98$ fm\\
  $R_2 = 0.49$, $\lambda_2 = 0.15$ (G)
& $R_2 = 0.13$, $\lambda_2 = 0.10$ (G)
& $R_0 = 0.22$\\
  $\lambda_1+\lambda_2=0.97$
& $\lambda_1+\lambda_2=0.81$
& $\lambda = 1.0\ (\alpha_{\tau} =0.56$)\\
  $c = 0.91$
& $c = 1.00$
& $c = 0.99$\\
  $\delta = 0.13$ fm
& $\alpha = -0.07$ fm$^2$
& \\ 
  $\varepsilon = -0.035$ fm$^2$
& $\beta = 0.65$ fm
& \\ 
  $\chi^2/{\rm ndf} = 209/191$
& $\chi^2/{\rm ndf} = 209/191$
& $\chi^2/{\rm ndf} = 240/192$\\
\hline
7 TeV\qquad $R$ (fm)\qquad & $R$ (fm) & $R$ (fm)\\
  $R_1 = 3.15$ fm, $\lambda_1 = 0.83$ (E)
& $R_1 = 3.42$ fm, $\lambda_1 = 0.75$ (E)
& $R = 3.46$ fm\\
  $R_2 = 0.53$ fm, $\lambda_2 = 0.12$ (G)
& $R_2 = 0.15$ fm, $\lambda_2 = 0.08$ (G)
& $R_0 = 0.22$ fm\\
  $\lambda_1+\lambda_2=0.95$
& $\lambda_1+\lambda_2=0.82$
& $\lambda = 1.0\ (\alpha_{\tau} = 0.56$)\\
  $c = 0.93$
& $c = 1.00$
& $c = 0.99$\\
  $\delta = 0.020$ fm
& $\alpha = -0.08$ fm$^2$
& \\ 
  $\varepsilon = -0.001$ fm$^2$
& $\beta = 0.72$ fm
& \\
  $\chi^2/{\rm ndf} = 208/191$
& $\chi^2/{\rm ndf} = 207/191$
& $\chi^2/{\rm ndf} = 289/192$\\
\hline
\hline
\multicolumn{3}{l}{Note: When ${\rm CF_{II}}\times {\rm LRC}_{(\delta)}$ is utilized in the analysis of the CMS BEC at 7 TeV, }\\
\multicolumn{3}{l}{the following estimated parameters are obtained~\cite{Biyajima:2019wcb}: }\\
\multicolumn{3}{c}{$R_1 = 3.88$ fm, $\lambda_1 = 0.84$ (E), $R_2 = 0.71$ fm, $\lambda_2 = 0.12$ (G), and $\chi^2 = 540$.}\\
\multicolumn{3}{l}{Adopting LRC$_{(\alpha,\ \beta,\ n=2)}$, we obtain a better $\chi^2$ value, as mentioned above.}\\
\hline
\end{tabular}
\end{table}

\section{\label{sec4}Analysis of OPAL BEC renormalized by Monte Carlo at $Z^0$-pole using Eq.~(\ref{eq7})}
Because the OPAL BEC at the $Z^0$-pole is not separable into ``2-jet'' and ``3-jet'' events, we apply Eq.~(\ref{eq7}) to this case. Because the OPAL BEC prefers Gaussian distributions, we choose a combination of G$+$G. 

\begin{figure}[H]
  \centering
  \includegraphics[width=0.48\columnwidth]{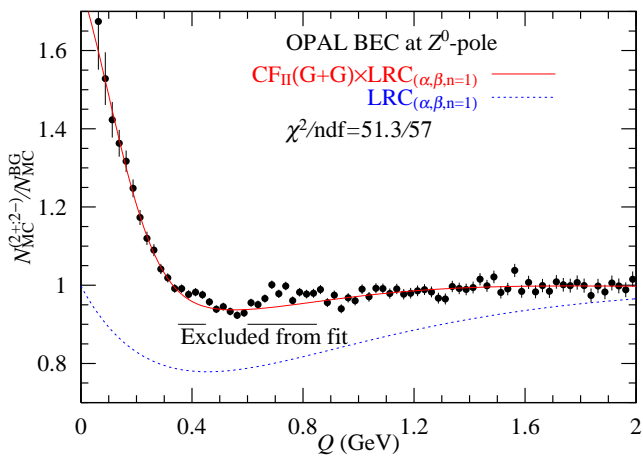}
  \caption{\label{fig8}Analysis of the OPAL BEC at the $Z^0$-pole using Eqs.~(\ref{eq4}) and (\ref{eq7}). }
\end{figure}

Our results are shown in Table~\ref{tab6} and Fig.~\ref{fig8}. In Table~\ref{tab6}, we observe that $R_2(=1.39$ fm) increases as $|\alpha|=2.41$ GeV$^{-1}$ increases; we conclude that these quantities are linked to each other.

\begin{table}[H]
\centering
\caption{\label{tab6}Analysis of the OPAL BEC at the $Z^0$-pole using Eqs.~(\ref{eq4}) and (\ref{eq7}). ``G'' indicates the Gaussian distribution.}
\vspace{1mm}
\renewcommand{\arraystretch}{1.2}
\begin{tabular}{cccccccc}
\hline
\lw{data} & \lw{$\!\! R_1$ (fm)(G)$\!\!$} & \lw{$\lambda_1$} & \lw{$\!\! R_2$ (fm)(G)$\!\!$} & \lw{$\lambda_2$} & \lw{$c$} & $\!\!\alpha$ (GeV$^{-1}$) & \lw{$\!\!\chi^2/$ndf}\\
&&&&&& $\!\!\beta$ (GeV$^{-1}$) & \\
\hline
\lw{$\!\! \dfrac{N^{(2+:\,2-)}}{N^{\rm BG}}\!\!$}
& \lw{0.16$\pm$0.02}
& \lw{0.34$\pm$0.06}
& \lw{1.39$\pm$0.19}
& \lw{0.33$\pm$0.07}
& \lw{0.95$\pm$0.02}
& $-$2.41$\pm$0.23
& \lw{74.8/57}\\
&&&&&& 2.33$\pm$0.06 &\\
\hline
\lw{$\!\! \dfrac{N_{\rm MC}^{(2+:\,2-)}}{N_{\rm MC}^{\rm BG}}\!\!$}
& \lw{0.14$\pm$0.04}
& \lw{0.23$\pm$0.01}
& \lw{0.91$\pm$0.04}
& \lw{0.54$\pm$0.06}
& \lw{1.00$\pm$0.04}
& $-$1.31$\pm$0.26
& \lw{51.3/57}\\
&&&&&& 2.20$\pm$0.19 &\\
\hline
\end{tabular}
\end{table}

\section{\label{sec5}Pion-pair density distributions in Euclidean space}
\begin{figure}[htbp]
  \centering
  \includegraphics[width=0.48\columnwidth]{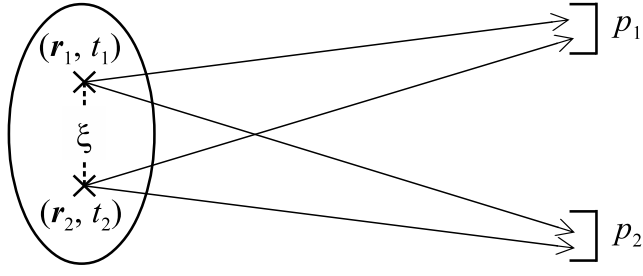}
  \caption{\label{fig9}Geometrical picture of the BEC. $\xi$ denotes the distance between two production points ($\bm r_1,\,t_1$) and ($\bm r_2,\,t_2$) in the Euclidean space-time. $p_1$ and $p_2$ are momenta of identical pion pair.}
\end{figure}

We are able to calculate the pion-pairs density distribution in Euclidean space via the Fourier analysis~\cite{Shimoda:1992gb,Levy:1950aa,Bateman:1954aa,Sneddon:1995aa}:
\begin{eqnarray}
\mbox{\Large $\rho$}_{\rm BE}(\xi,\,R) = \frac 1{(2\pi)^2\xi} \int_0^{\infty} Q^2E_{\rm BE}(Q,\,R)J_1(Q\xi)dQ
\label{eq9},
\end{eqnarray}
where $\xi = \sqrt{x^2+y^2+z^2+(ct)^2}$, and $J_1(Q\xi)$ is the modified Bessel function. The variable $\xi$ is displayed in Fig.~\ref{fig9}.

For the LRC, by substituting the expression $({\rm LRC}-1.0)=\alpha Q^n e^{-\beta Q}$ into Eq.~(\ref{eq9}), we obtain
\begin{eqnarray}
\mbox{\Large $\rho$}_{\rm (LRC-1)}(\xi,\,\alpha,\,\beta,\,n=2)&=& \frac 1{(2\pi)^2\xi} \int_0^{\infty} Q^2 [{\rm LRC}(Q,\,\alpha,\,\beta,\,n=2)-1.0]J_1(Q\xi)dQ\nonumber\\
&=& \frac{\alpha}{(2\pi)^2\xi}\frac{\Gamma(6)}{(\beta^2+\xi^2)^{5/2}}{\rm P}_4^{-1}\left(\frac{\beta}{\sqrt{\beta^2+\xi^2}}\right),
\label{eq10}
\end{eqnarray}
where $P_4^{-1}(x)$ is the associated Legendre function and $\Gamma(x)$ is the gamma function. 

\begin{figure}[H]
  \centering
  \includegraphics[width=0.48\columnwidth]{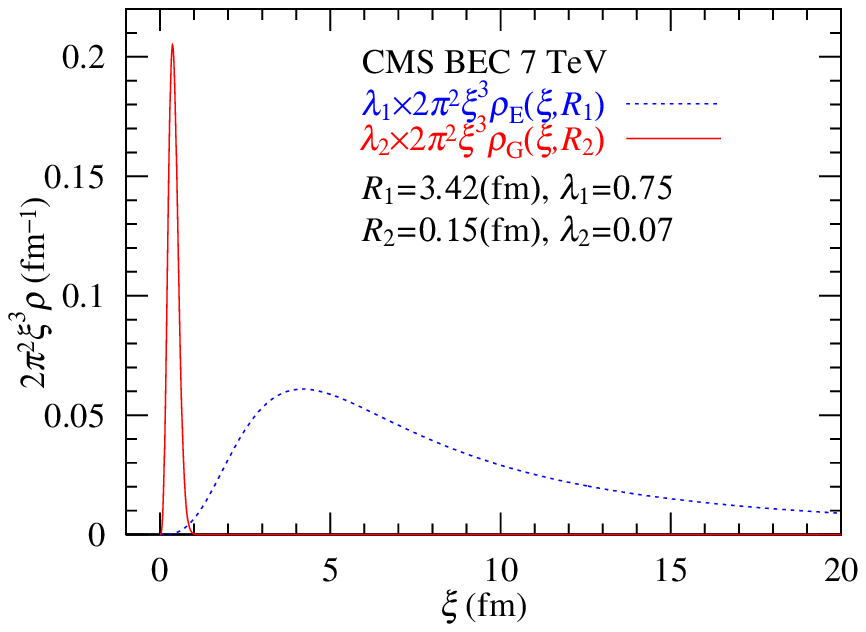}
  \includegraphics[width=0.48\columnwidth]{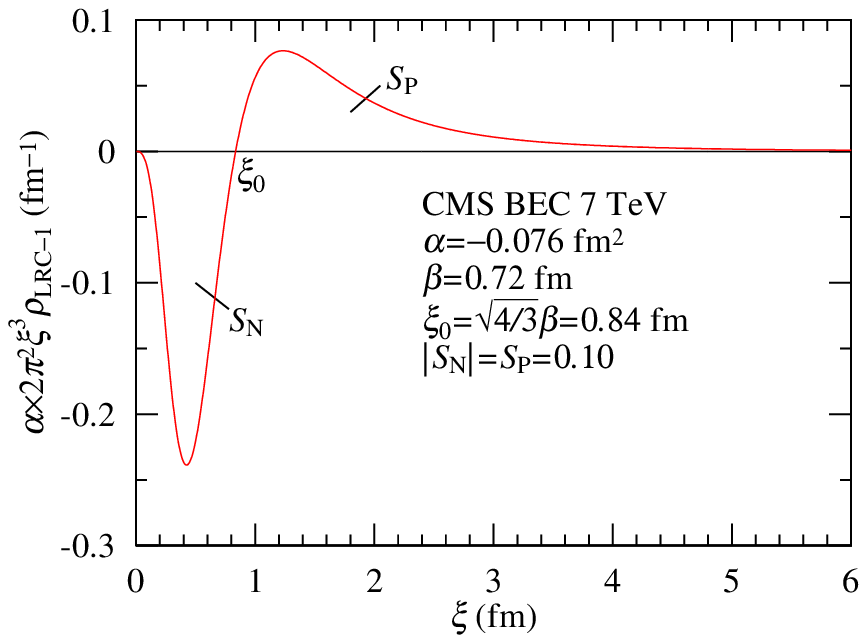}\\
  \includegraphics[width=0.65\columnwidth]{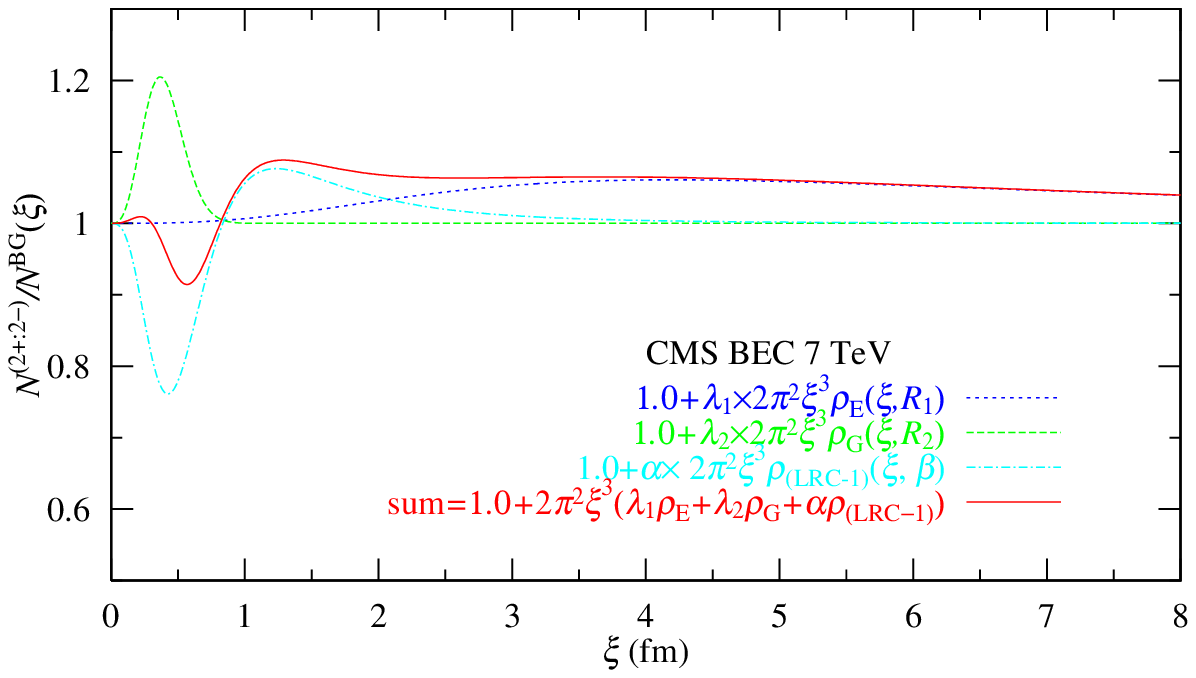}
  \caption{\label{fig10}Pion-pairs density distributions of the CMS BEC at 0.9 and 7 TeV in four-dimensional Euclidean space, where $\xi$ denotes the distance between two pion-production points. The Contributions of the crossed terms ($\rho_{\rm G}\times \rho_{\rm (LRC-1)}$ and $\rho_{\rm E}\times \rho_{\rm (LRC-1)}$) are invisible. Maximal points in the distributions are shifted by the phase space $2\pi^2\xi^3$: $R_{\rm G}\to \sqrt{6}R_{\rm G}$ and $R_{\rm E}\to \sqrt{3/2}R_{\rm E}$.}
\end{figure}

\begin{figure}[H]
  \centering
  \includegraphics[width=0.48\columnwidth]{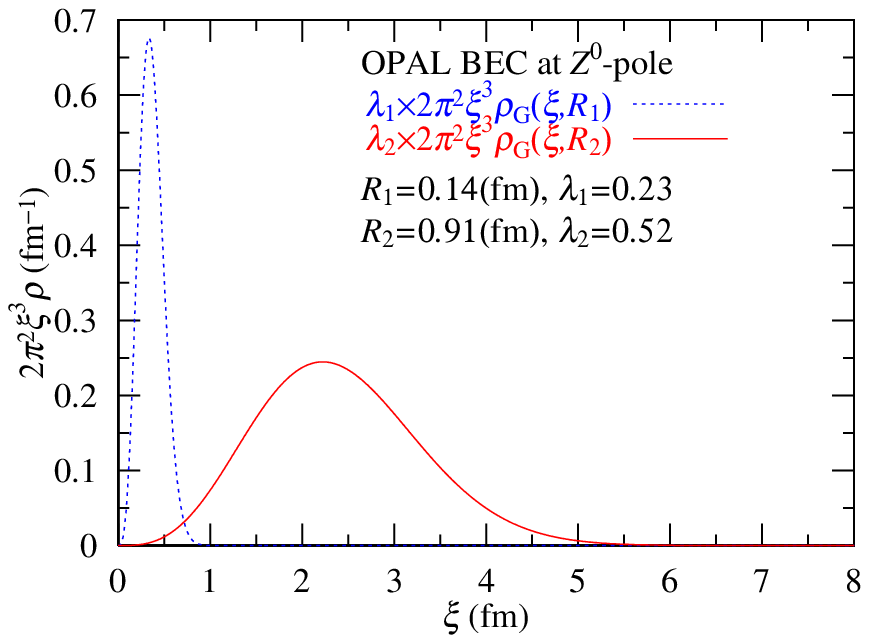}
  \includegraphics[width=0.48\columnwidth]{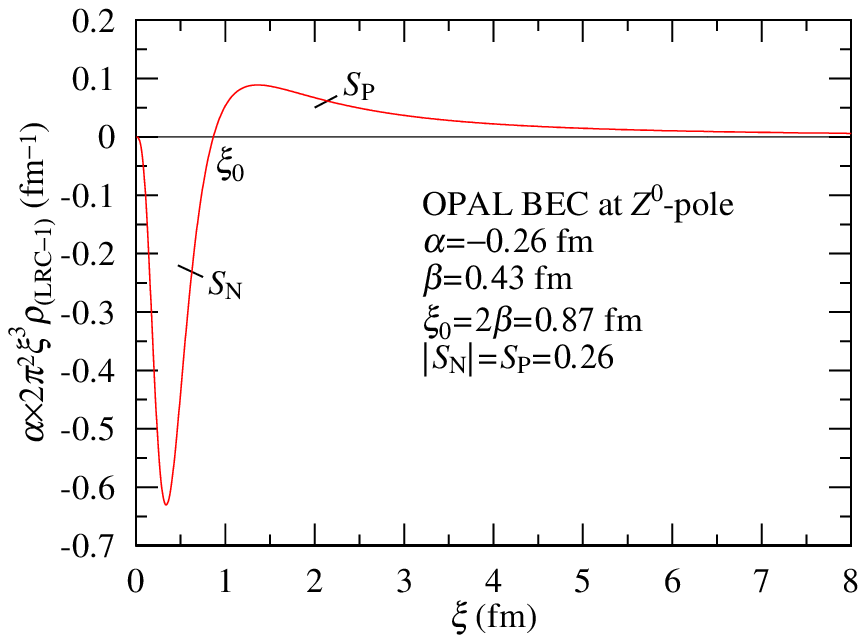}\\
  \includegraphics[width=0.65\columnwidth]{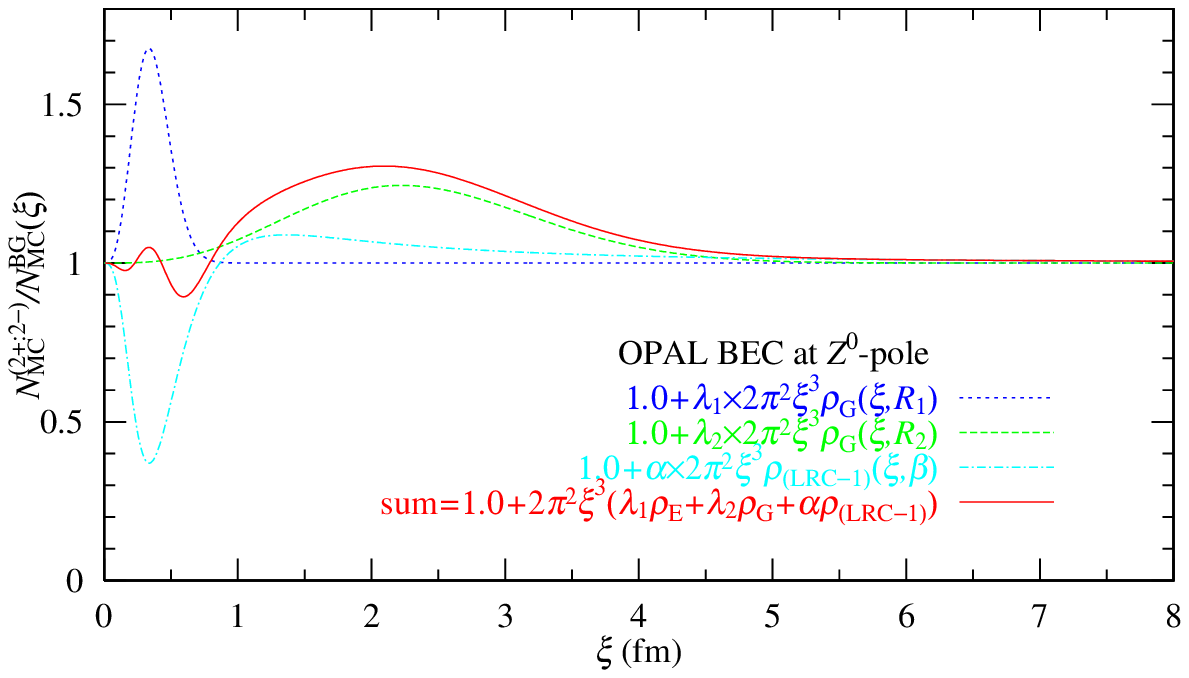}
  \caption{\label{fig11}Pion-pairs density distributions of the OPAL BEC at the $Z^0$-pole in four-dimensional Euclidean space, where $\xi$ denotes the distance between two pion-production points. The contributions of crossed terms ($\rho_{\rm G}\times \rho_{\rm (LRC-1)}$ and $\rho_{\rm E}\times \rho_{\rm (LRC-1)}$) are invisible. Maximal points in the distributions are shifted by the phase space $2\pi^2\xi^3$: $R_{\rm G}\to \sqrt{6}R_{\rm G}$.}
\end{figure}

For $n=1$, we have the following formula:
\begin{eqnarray}
\mbox{\Large $\rho$}_{\rm (LRC-1)}(\xi,\,\alpha,\,\beta,\,n=1)=\frac{\alpha}{(2\pi)^2\xi}\frac{\Gamma(5)}{(\beta^2+\xi^2)^{2}}{\rm P}_3^{-1}\left(\frac{\beta}{\sqrt{\beta^2+\xi^2}}\right).
\label{eq11}
\end{eqnarray}
The pion-pairs density distributions are shown in Table~\ref{tab7} and Figs.~\ref{fig10} and \ref{fig11}. The suffixes E and G indicate the exponential functions and Gaussian distributions, respectively. The contributions of the Gaussian distributions may contain contamination between different hadron-pairs, resonances, and/or energy conservation among produced hadrons.

\begin{table}[H]
\centering
\caption{\label{tab7}Correlation functions, source functions, and pion-pairs density distribution.} 
\vspace{1mm}
\renewcommand{\arraystretch}{2.0}
\begin{tabular}{c|c|c}
\hline
  $E_{\rm BE}(R,\,Q)$ & $\rho(\xi)$ & number of pairs density distributions\\
\hline
$\exp(-R^2Q^2)$
& $\dfrac 1{16\pi^2R^4}\exp\left(-\dfrac{\xi^2}{4R^2}\right)$
& \lw{$2\pi^2\xi^3 \rho(\xi)$}\\
$\exp(-RQ)$
& $\dfrac 3{4\pi^2R^4}\dfrac 1{(1+(\xi/R)^2)^{5/2}}$
& ($2\pi^2\xi^3$ : phase space)\\
\hline
\end{tabular}
\end{table}

\section{\label{sec6}Concluding remarks and discussions}
The following has been concluded:

\paragraph{C1)} In many analyses of BEC, the phenomenological forms of LRC have been adopted. In this paper,  we propose an analytic form, i.e., Eq.~(\ref{eq4}), which is probably reproducing producing Eqs.~(\ref{eq2}) and (\ref{eq3}).

\paragraph{C2)} As seen in Table~\ref{tab3}, four kinds of analyses for the OPAL BEC at the $Z^0$-pole show almost the same values of $R$s(G) ($\sim$0.9 fm) and $\chi^2$ ($\sim$110). This means that there is no large, negative ($\alpha$) contribution at $n=1$. 

\paragraph{C3)} When analyzing the L3 and CMS BECs, we can compare our results obtained using CF$_{\rm I}$ and CF$_{\rm II}$ with those using Eq.~(\ref{eq4}) with the same quantities estimated using the $\tau$-model. As seen in Tables~\ref{tab4} and \ref{tab5}, it can be said that they are almost the same, provided that Eq.~(\ref{eq4}) is utilized for CF$_{\rm I}$ and CF$_{\rm II}$. On the contrary, the results from CF$_{\rm II}\times$LRC$_{(\delta)}$ are improved using LRC$_{(\alpha,\ \beta,\ n=2)}$. See the right column of Table~\ref{tab5}.

\paragraph{C4)} As seen in Tables~\ref{tab2}, \ref{tab4}, \ref{tab5}, and \ref{tab7}, the exchange functions (i.e., the Gaussian distribution and the exponential function) and the power number ($n$) of the ${\rm LRC}_{(\alpha,\, \beta,\, n)}$ are related to one another (see Table~\ref{tab8}). In other words, the deformations of exchange functions due to LRC$_{(\alpha,\ \beta,\ n=2)}$ are necessary for analysis of BECs.

\paragraph{C5)} As seen in Figs.~\ref{fig10} and \ref{fig11}, the pion-pairs density distributions in the regions with $\xi\le 1$ fm are very similar to each other. In the region above 1 fm, we see a Gaussian distribution for the $Z^0$-pole and an inverse power law with $s=5/2$ for $pp$ collisions at the LHC. This suggests that the interaction region in the $pp$ collisions at the LHC is larger than that at the $Z^0$-pole.

\paragraph{D1)} To describe the distributions in Minkowski space~\cite{Bogoliubov:1980aa}, we need distributions on energy differences ($\Delta E =(p_{10}-p_{20})$). 

\paragraph{D2)} A study of the Fourier transform in the Levy stochastic process~\cite{Zolotarev:1981aa,Wilk:1999dr} is necessary for advanced investigations.

\paragraph{D3)} For presently unclear reasons, the OPAL BEC preferred the Gaussian distribution to the exponential function, whereas the L3 BEC with separable data (2- and 3-jet cases) takes the exponential function.

\paragraph{D4)} As seen in Table~\ref{tab5}, $R_2=0.15$ fm when estimated using ${\rm CF_{II}}\times {\rm LRC}_{(\alpha,\ \beta,\ n=2)}$ and $R_2=0.22$ fm when estimated using the $\tau$-model, which are shown respectively. It is not clear why these values are so similar.

\begin{table}[H]
\centering
\caption{\label{tab8}Empirical relationship between exchange the functions and ${\rm LRC}_{(\alpha,\, \beta,\, n)}$.}
\vspace{1mm}
\begin{tabular}{cccc}
\hline
$E_{\rm BE}(R,\,Q)$ & & ${\rm LRC}_{(\alpha,\, \beta,\, n)}$ & data\\
\hline
Gaussian distribution & $\!\!\!\longleftrightarrow\!\!\!$ & $n=1$ & OPAL\\
Exponential function & $\!\!\!\longleftrightarrow\!\!\!$ & $n=2$ & L3, CMS\\
\hline
\end{tabular}
\end{table}

\noindent
{\it Acknowledgments.} We are thankful to the organizer of 2020 Zimanyi Winter School and various comments presented there. Concerning L3 BEC data, we are indebted to W.~J.~Metzger and M.~Csanad for their kindness. M.~Biyajima thanks his colleagues at the Department of Physics of Shinshu University for their kindness.


\end{document}